\documentclass[11pt]{article}
\usepackage{geometry}
\usepackage{graphicx}
\usepackage{tensor}
\usepackage{url}
\usepackage{bm}
\usepackage{bbm}
\usepackage{amsthm}
\usepackage{amsfonts}
\usepackage{tabto}
\usepackage[english]{babel}
\usepackage[T1]{fontenc}
\usepackage[utf8]{inputenc}
\usepackage{xcolor}
\usepackage{tensor}
\usepackage{commath}
\usepackage{mathtools}
\usepackage{braket}
\usepackage{physics}
\usepackage{float}
\usepackage{subfig}
\usepackage[hidelinks]{hyperref}
\usepackage{cleveref}
\usepackage{authblk}
\usepackage{amsmath}
\usepackage{amssymb}
\usepackage{tabu}

\usepackage{soul}

\newcommand{\ebd}{\vcentcolon =}
\newcommand{\ad}{\text{ad}}

\makeatletter
\newcommand{\fmarki}{*}
\newcommand{\fmarkii}{\ensuremath{\dagger}}
\newcommand{\fmarkiii}{\ensuremath{\ddagger}}

\def\@fnsymbol#1{{\ifcase#1\or \fmarki\or \fmarkii\or \fmarkiii \else\@ctrerr\fi}}
\makeatother

\renewcommand{\fmarki}{$\star$}
\renewcommand{\fmarkii}{$\natural$}
\renewcommand{\fmarkiii}{$\mathsection$}

\begin{document}

\title{{\bf Fundamental decoherence from quantum spacetime}}
\date{}
\author[]{Michele Arzano\thanks{michele.arzano@na.infn.it} \and Vittorio D'Esposito\thanks{vittorio.desposito@unina.it} \and Giulia Gubitosi\thanks{giulia.gubitosi@unina.it}}
\affil[]{Dipartimento di Fisica “E. Pancini”, Università di Napoli Federico II \\ INFN sezione di Napoli,
Complesso Universitario di Monte S. Angelo Edificio 6, via Cintia, 80126 Napoli, Italy}

\maketitle

\begin{abstract}
We show that quantum properties of spacetime, encoded by noncommutativity at the Planck scale, lead to a generalized time evolution of quantum systems in which pure states can evolve into mixed states. Specifically, a decoherence mechanism is obtained in the form of a Lindblad-like time evolution for the density operator when the action of time translations generator is deformed by the effects of spacetime noncommutativity. The decoherence time for the evolution of a free particle is used to show that the Planck mass is the maximum allowed mass for elementary quantum systems.  
\end{abstract}

\section{Introduction}

The emergence of the classical macroscopic world from the microscopic quantum realm is a long-standing open problem, still  heavily debated in the physics community and subject to theoretical and experimental research \cite{Schlosshauer:2003zy, Giulini:2003}.
 Several different mechanisms causing  decoherence of quantum systems have been considered. They are not  mutually exclusive, and their relative relevance might vary depending on the specific physical setup. The most studied and  best understood possibility relies on the interaction of quantum systems with an environment \cite{Schlosshauer:2003zy, Giulini:2003}. The resulting entanglement between the respective degrees of freedom leads to the emergence of  classical properties for the quantum system. 
Growing theoretical evidence  \cite{Bassi:2017szd} indicates that  gravity might contribute to environmental decoherence, if one accounts for the entanglement between the gravitational degrees of freedom (either classical or quantum) and those of the quantum system \cite{Kiefer:1987ft, Ellis:1988uk, Joos1985, Blencowe:2012mp, Garay:1999cy, Mavromatos:2004sz}. Gravity might also induce decoherence via different mechanisms. In fact, an enticing possibility is that classical or quantum gravity effects might cause departures from standard quantum theory, as proposed in the pioneering works \cite{Diosi1984, Percival:1995an}. This type of decoherence is called fundamental, since it would affect the quantum system independently of its environment. As opposed to environmental decoherence, where information on quantum coherence is delocalized in the environment, in the case of fundamental decoherence such information is lost, since there is no environment to which it could have been transferred.

Fundamental decoherence opens up the possibility that quantum gravity might be tested far away from the Planck scale, for example in nonrelativistic quantum systems and in table-top experiments \cite{Amelino-Camelia:2005xme, Carney:2018ofe, Pfister2016}. After almost a century of intense theoretical research which did not lead to a solid understanding of the interplay between the quantum realm and gravity, this might provide a turning point.

The possibility that the interplay between quantum and gravitational effects  can generate decoherence of quantum systems emerged already in the earliest investigations on quantum gravity. 
The original motivation was provided by Hawking's proposal \cite{Hawking1976} that the evaporation of a black hole evolves pure states into mixed states. 
This is at the basis of the so-called black hole information paradox, which is still nowadays at the center of a heated debate aimed at establishing whether the evolution of the quantum states of a black hole and the emitted radiation is unitary  \cite{Banks:1983by, Susskind:1993if, Kiem:1995iy, Kiefer:2001wn,  Hawking_2005,  PhysRevLett.110.101301,  Unruh_2017}. 
Further motivation comes from the need to explain the evolution of  quantum vacuum fluctuations generated in the primordial universe so that they turn into the classical large-scale structures observed today. How this quantum-to-classical transition can occur is one of the most pressing open problems in cosmology and has been the subject of many studies \cite{Brandenberger:1992sr, Perez:2005gh, Kiefer:2008ku, Sudarsky:2009za, Gubitosi:2017zoj, Martin:2018zbe}.
It is then clear that establishing the role of gravity in the generation of decoherence would have far-fetching implications for a number of issues that lie at the  foundations of several areas of physics research.

Several quantum gravity-induced decoherence mechanisms proposed in the literature relied on heuristic analyses. In particular, putative properties of spacetime
induced by quantum gravity effects were introduced in a rather ad-hoc manner in the derivation of decoherence,  based on general ideas about gravity at the Planck scale, such as discretization \cite{Gambini:2004de}, foaminess \cite{Garay:1999cy, Mavromatos:2004sz}, stochasticity \cite{Percival:1995an, Anastopoulos:2013zya, Breuer:2008rh}, fundamental uncertainties in length measurements \cite{Petruzziello2021, Ng:1993jb}.

In this work, we provide the first rigorous derivation of decoherence within a full-fledged model of quantum spacetime, namely spacetime noncommutativity \cite{Connes:1994, Szabo:2001kg}\footnote{ "Quantum spacetime" refers to a spacetime whose properties are modified because of  quantum-gravitational effects, see e.g. the review  \cite{Amelino-Camelia:2008aez}. While in some models this might be seen as an abuse of notation, the case of noncommutative spacetime is perhaps the one where the terminology is the most accurate, since noncommuting spacetime coordinates can be described using the theory of operators on a Hilbert space, see e.g. \cite{Doplicher:1994zv, Lizzi:2018qaf}.}. Spacetime noncommutativity was proposed already in the early days of quantum mechanics, most notably by Heisenberg \cite{HeisenbergLetter} and Snyder \cite{Snyder:1946qz}, as a means to introduce an effective ultraviolet cutoff and control the divergences in quantum field theory without breaking Lorentz symmetries. The idea was later revived in the 90's, when it was realized that it provides a framework to describe the fundamental limitations emerging when trying to localize an event with a spatial accuracy comparable with the Planck length, due to the interplay of quantum and gravitational effects \cite{Doplicher:1994zv, Ahluwalia:1993dd}. Over the past twenty years, the relevance of spacetime noncommutativity for quantum gravity research has been  established on solid grounds: on the one hand it emerges as an effective  description of spacetime in some regimes of string theory \cite{Veneziano:1986zf, Gross:1987ar, Amati:1988tn} and loop quantum gravity \cite{Amelino-Camelia:2003ezw, Freidel:2005me, Amelino-Camelia:2016gfx}, and on the other hand it  provides a source of rich phenomenology \cite{Amelino-Camelia:2008aez, Addazi2022}.

One of the most studied classes of noncommutative spacetime models is the so-called $\kappa$-Minkowski spacetime \cite{Majid1994}, characterized by noncommutativity between the time and spatial coordinates: 
\begin{equation}
    \comm{x^0}{x^i} = \frac{i}{\kappa}x^i \,, \quad\comm{x^i}{x^j} = 0\,.\label{eq:kM}
\end{equation}
Here, $\kappa^{-1}$ is a length scale, usually identified with the Planck length $\ell_P$ (unless otherwise stated, we use units such that $c=\hbar=1$). In the following we will take $\kappa>0$. Notice that taking the opposite sign for $\kappa$ amounts to considering a different physical model, which would lead to different implications concerning decoherence.

For our purposes, the most relevant feature of \eqref{eq:kM} is the fact that its relativistic symmetries are described by a quantum deformation of the Poincar\'e symmetries, encoded by the $\kappa$-Poincar\'e Hopf algebra \cite{ Lukierski:1991pn,Lukierski:1993wxa}. In particular, the time evolution operator turns out to have a deformed action on the density operator and the ensuing nonstandard time evolution generates decoherence. Ultimately, decoherence emerges because the time evolution of quantum systems must not spoil the relativistic invariance of the noncommutative spacetime. Therefore, the decoherence we derive within this model is fundamental, in the sense described above: there is no actual environment or interaction inducing decoherence; rather, this is due to the properties of spacetime itself and as such no quantum system can avoid it.

The paper is organized as follows. In  section \ref{sec:Deformations} we introduce the novel structures that are needed to describe quantum deformations of spacetime symmetries. We discuss how these deformations emerge because of spacetime noncommutativity and we present the deformed time evolution operator. In section  \ref{sec:Evolution} we use the  time evolution operator to define a Linblad-like equation \cite{Lindblad1976} describing the time evolution of the density operator of a generic quantum system. We show that this time evolution generally implies an increase of the linear entropy,  signalling decoherence. Specializing to the case of a free particle, 
we obtain the decoherence time for a quantum system in the momentum basis. The decoherence rate is suppressed by the quantum deformation scale $\kappa$ and is amplified by increasing values of momenta: the efficiency of the decoherence process grows with the characteristic energy of the quantum system. The quadratic momentum dependence of the decoherence time allows us to obtain intrinsic fundamental constraints on the mass of elementary quantum systems, which cannot exceed the quantum deformation scale $\kappa$.  This shows that fundamental decoherence puts an ultimate bound on the scales at which quantum features of nature can be observed. Had the momentum dependence been different, we would not have been able to derive such absolute bound. This  bound is thus a genuine implication of quantum spacetime properties.
Further theoretical and observational implications are discussed in the outlook section.

\section{Deformations of spacetime symmetries}
\label{sec:Deformations}

In standard quantum mechanics, for a quantum system with symmetry under translations we can define a basis of momentum eigenstates on the Hilbert space of the system $\mathcal{H}$,
\begin{equation}
P_i \ket{k} = k_i\ket{k}\,.
\end{equation}
For a free particle, these eigenstates can be represented as plane waves, $\ip{x}{k} = e^{i k_\mu x^\mu}$.
This representation can be extended to the dual space $\mathcal{H}^* $ spanned by bras: 
\begin{equation}\label{dualrep}
P_{i}\bra{k} = - k_{i} \bra{k}\,.
\end{equation}
The representation \eqref{dualrep} can be written in terms of the {\it antipode map} $S(P_i)$ connecting the left and right action of $P_i$ on dual states:
\begin{equation}\label{antipode}
    P_i \bra{k} = - k_i \bra{k} =  \bra{k} (- k_i) \equiv \bra{k} S(P_i)\,,
\end{equation}
where 
\begin{equation}\label{antipi}
 S(P_i) = -P_i\,.   
\end{equation}
 Multi-particle states are given by tensor products of the Hilbert space $\mathcal{H}$. The representation of $P_{i}$ on a two-particle state is given by the familiar Leibniz rule, which can be written in terms of a {\it coproduct} map 
\begin{equation}\label{coprpi}
 \Delta P_{i} = P_{i} \otimes \mathbbm{1} +\mathbbm{1}\otimes P_{i}\,,
\end{equation}
such that
\begin{eqnarray}
    \Delta P_{i} \qty(\ket{g} \otimes \ket{h}) &=&\qty( P_i \ket{g}) \otimes \ket{h}+\ket{g} \otimes \qty(P_i \ket{h})\nonumber\\
    &=& \qty(g_i+h_i) (\ket{g} \otimes \ket{h}) \,.
\end{eqnarray}

For a non-commutative space-time defined by commutators like \eqref{eq:kM} which close a Lie algebra, plane waves are generalized to noncommutative ordered exponentials $:~e^{i k_\mu x^\mu}:$ \cite{Amelino-Camelia:1999jfz}. The parameters appearing in these exponentials are  coordinates of a non-abelian Lie group rather than an ordinary vector space \cite{Kowalski-Glikman:2013rxa, Arzano:2022ewc}.
Quantum states in the momentum representation are then labelled by coordinates on the group $\ket{k(g)}$, such that the action of translation generators is given by
\begin{equation}
P_{i} \ket{k(g)} = k_{i}(g)\ket{k(g)}\,.
\end{equation}
The homomorphism property of coordinate functions tells us that momentum eigenvalues $k_{i}(g)$ obey a deformed, non-abelian, composition law, 
\begin{equation}
   k_{i}(g)\oplus k_{i}(h)\equiv k_{i}(gh)\neq k_{i}(h)\oplus k_{i}(g)\equiv k_{i}(hg)\,, 
\end{equation}
and this can be used to define the inverse group element $k_{i}(g^{-1})$, such that
\begin{equation}
   k_{i}(g)\oplus k_{i}(g^{-1})=0\,.
   \end{equation}
As a consequence, the antipode and coproduct maps defined above are deformed, thus encoding a deformation of translational symmetries.
For the analogue of the dual representation \eqref{dualrep}-\eqref{antipode} the antipode gives the coordinates of the inverse momentum group element,
\begin{equation}
P_{i} \bra{k(g)} = k_{i}(g^{-1})\bra{k(g)}= \bra{k(g)}\, S(P_{i}) \,,
\end{equation}
while the coproduct allows us to establish the total momentum carried by a two-particle state,
\begin{equation}
    \Delta P_{i} (\ket{g} \otimes \ket{h}) =  k_{i}(gh) (\ket{g} \otimes \ket{h}) \,.
\end{equation}
For the specific model we are considering, in which spacetime noncommutativity is determined by the commutators \eqref{eq:kM}, the coproduct and antipodes for translation generators read
\begin{eqnarray}
     \Delta P_0 &=& P_0\otimes \Pi_0 +\Pi_0^{-1}\otimes P_0 + \frac{1}{\kappa} P_i\Pi_0^{-1}\otimes P^i \,,\nonumber \\
    \Delta P_i &=& P_i\otimes \mathbbm{1} +\mathbbm{1}\otimes P_i\,, \; \; \\
     S(P_0) &=& -P_0 + \frac{1}{\kappa} \pmb{P}^2\, \Pi_0^{-1}\,, \; \;     S(P_i) = -P_i\, \Pi_0^{-1},\nonumber \\
\end{eqnarray}
where
\begin{eqnarray}
    \Pi_0 &=& \frac{1}{\kappa}P_0 + \sqrt{1-\frac{1}{\kappa^2}P^\mu P_\mu}\,, \nonumber \\ \Pi_0^{-1} &=& \frac{\sqrt{1-\frac{1}{\kappa^2}P^\mu P_\mu}-\frac{1}{\kappa}P_0}{1-\frac{1}{\kappa^2}\pmb{P}^2}\,, \nonumber\\
    P^\mu P_\mu&=&-P_0^2+\pmb{P}^2 = -P_0^2+P_iP^i\,. \nonumber
\end{eqnarray}

This particular realization of the deformed generators of relativistic translations is the translation sector of the so-called {\it classical basis} \cite{Borowiec2009} of the $\kappa$-Poincar\'e algebra. In such model only coproducts and antipodes are deformed, while the commutators between the generators of space and time translation, boosts and rotations, i.e. the Poincar\'e Lie algebra, remain undeformed.

 Since the symmetry group relevant for nonrelativistic quantum mechanics is the Galilei group rather than the Poincar\'e group, we consider a nonrelativistic limit of the  $\kappa$-Poincar\'e algebra, namely, the $\kappa$-Galilei algebra. 
 This limit is obtained by means of a standard Hopf-algebra contraction  procedure \cite{Ballesteros:2020uxp, Ballesteros:2021dob}. The speed of light $c$ is used to rescale both the relevant generators, namely, space translations $P_i$ and boosts $N_i$, and the quantum deformation parameter,
\begin{equation}
    {P_i} \mapsto c^{-1}{P_i} \,, \; \; {N_i} \mapsto c^{-1}{N_i} \,, \; \; \kappa \mapsto c^{-2}\kappa\,.
    \label{contraction}
\end{equation}
After taking the limit $c\rightarrow \infty$, the commutation relations of the Poincar\'e Lie algebra give the standard Galilei algebra, while the following nontrivial structures in the coalgebra sector are obtained: 
\begin{equation}
\begin{array}{ll}
    \Delta P_0 &= P_0\otimes \mathbbm{1} +\mathbbm{1}\otimes P_0 + \frac{1}{\kappa} P_i\otimes P^i\,,  \\
     S(P_0) &= -P_0 + \frac{1}{\kappa} \pmb{P}^2\,.
    \label{abstract coalgebra galilei}
    \end{array}
\end{equation}

In order to study the evolution of quantum systems,  we use the familiar $G\mapsto -i \,G$ procedure  to map the abstract  elements $G$ of the $\kappa$-Galilei algebra into Hermitian operators that can be represented on the algebra of operators of a quantum system. We thus obtain the following 
coalgebra relations: 
\begin{equation}
    \Delta P_0 = P_0\otimes \mathbbm{1} +\mathbbm{1}\otimes P_0 - \frac{i}{\kappa} P_i\otimes P^i\,,\quad  S(P_0) = -P_0 - \frac{i}{\kappa} \pmb{P}^2\,.
    \label{coalgebra galilei}
\end{equation}

Notice how under the same $G\mapsto -i \,G$ transformation the nontrivial coproducts and antipodes for the translation generators of the $\kappa$-Poincar\'e algebra  become
\begin{equation}\label{coalgebra poincare}
\begin{array}{ll}
     \Delta P_0 &= P_0\otimes \mathbbm{1} +\mathbbm{1}\otimes P_0 - \frac{i}{\kappa} P_i\otimes P^i\,,  \\  S(P_0) &= -P_0 - \frac{i}{\kappa} \pmb{P}^2 \,,\\
     \Delta P_i &= P_i\otimes\mathbbm{1} + \mathbbm{1} \otimes P_i - \frac{i}{\kappa} P_i \otimes P_0\,,  \\ S(P_i) &= -P_i - \frac{i}{\kappa} P_iP_0\,, 
    \end{array}
\end{equation}
 at first order in $\kappa^{-1}$. In particular, we notice that the coalgebra properties of the time translation generator $P_0$ are the same as those found for the time generator of the $\kappa$-Galilei algebra \eqref{coalgebra galilei}. As we will see in the following section, this correspondence guarantees that the same decoherence produced by the $\kappa$-deformation of the Galilei algebra is also produced by the $\kappa$-deformation of the Poincar\'e algebra to the first order in $\kappa^{-1}$.

\section{Generalized quantum evolution from deformed symmetries}\label{sec:Evolution}

The deformation of relativistic symmetries described in the previous section affects the evolution of quantum systems and induces decoherence. This can be shown by considering the action of the generators of translations on the algebra of operators of a quantum system $\mathcal{A}_{\mathcal{H}}$, with $\mathcal{H}$ being the Hilbert space of the quantum system itself.

Usually such action is given in terms of the standard commutator $\comm{\boldsymbol{\cdot}}{\boldsymbol{\cdot}}$, namely, via the adjoint action of the algebra of symmetry generators on $\mathcal{A}_{\mathcal{H}}$. When the Lie algebra of symmetry generators is generalized to a nontrivial Hopf algebra, the deformed coproducts and antipodes lead to a generalization of this action given by the {\it quantum adjoint action}  \cite{Ruegg:1994bk}.  This is defined  by \cite{Arzano:2014cya}
\begin{equation}
    \ad_{\,\boldsymbol{\cdot}} \,\boldsymbol{\cdot} \vcentcolon \mathcal{R}_G\times \mathcal{A}_{\mathcal{H}} \rightarrow \mathcal{A}_{\mathcal{H}}\,, \quad \ad_G \,O\ebd \left(id \otimes S\right)\Delta G \diamond O\,, \label{eq:adjAction}
\end{equation}
where $\mathcal{R}_G$ is the vector space spanned by the representatives of the generators of the symmetry algebra, $G\in \mathcal{R}_G$ and the $\diamond$ operator is defined by
\begin{equation}
    (a\otimes b) \diamond O \ebd a\, O\, b\,.
\end{equation}

This generalization of the adjoint action reduces to the standard one when the coproduct and antipode are the standard ones like e.g. in \eqref{antipi} and \eqref{coprpi}. 
In fact, given $A \in \mathcal{A}_{\mathcal{H}}$ and $G$ with trivial coproduct $\Delta G=G\otimes \mathbbm{1}+\mathbbm{1}\otimes G$ and antipode $S(G)=-G$, one finds
\begin{equation}
    \ad_G \, A = (G\otimes\mathbbm{1} -\mathbbm{1}\otimes G) \diamond A = \,\comm{G}{A}\,.\label{eq:comm}
\end{equation}

 In ordinary quantum mechanics, the evolution of the density operator  $\rho$ is given by the (standard) adjoint action \eqref{eq:comm} of the Hamiltonian, generator of time translations, on the density operator itself:
\begin{equation}
    i\,\partial_t \rho = \ad_H \, \rho = \, \comm{H}{\rho}\,. \label{eq:standardevolution}
\end{equation}
This is the standard Liouville--von Neumann equation. 

The generalised adjoint action \eqref{eq:adjAction} can be used to define the evolution equation for the density operator of a quantum system when the symmetry generators are those of the $\kappa$-Galilei Hopf algebra described in the previous section. Assuming that the time evolution is given by the time translation generator $P_0$, we define
\begin{equation}
       i\, \partial_t \rho \ebd \frac{1}{2}\qty{\ad_{P_0} (\rho) - \qty[\ad_{P_0}(\rho)]^\dagger}\,.
    \label{evolution}
\end{equation}
This reproduces the standard evolution equation \eqref{eq:standardevolution} for ordinary time translations with standard antipode and coproduct. Notice that we considered the combination of $\ad_{P_0} (\rho)$ and $\qty[\ad_{P_0}(\rho)]^\dagger$ in \eqref{evolution} so that the evolution preserves the Hermiticity of $\rho$.

The evolution equation \eqref{evolution} encodes an equation of Lindblad type \cite{Lindblad1976, Gorini:1975nb}:
\begin{equation}
    \partial_t \rho = -i \, \comm{H}{\rho} - \frac{1}{2}g^{mn}\left(Q_mQ_n\,\rho +\rho\, Q_mQ_n-2\, Q_m \,\rho\, Q_n\right)\,, \label{eq:Lindblad}
\end{equation}
where $Q_n$ form a basis of Hermitian operators. Indeed, after some algebra one can show that 
\begin{equation}
    \partial_t \rho = -i \, \comm{P_0}{\rho} - \frac{1}{2\kappa}\left(\pmb{P}^2 \rho + \rho\, \pmb{P}^2 - 2\,P_i\,\rho P^i\,\right)\,,
    \label{Lindblad Galilei}
\end{equation}
which reproduces \eqref{eq:Lindblad} for $g^{mn} = \kappa^{-1}\delta^{mn}$, $H=P_0$ and $Q_m=P_m$. The fact that the evolution equation \eqref{Lindblad Galilei} takes the form \eqref{eq:Lindblad}, with a   real and positive-definite  matrix $g^{mn}$,  guarantees that the operator $\rho$ is a good density operator at any time, namely, \eqref{Lindblad Galilei} preserves the positivity and the Hermiticity\footnote{The link between noncommutative deformations of relativistic symmetries and  generalized time evolution for quantum systems was noticed by one of the authors a few years ago \cite{Arzano:2014cya}. However, in \cite{Arzano:2014cya} the time evolution is not of the Lindblad type and conservation of the positivity of $\rho$ is not guaranteed.} of $\rho$. 

Equation \eqref{Lindblad Galilei}   describes the evolution of the density operator of a quantum system living on a noncommutative spacetime and whose symmetries are those of a quantum deformation of the Galilei algebra: the $\kappa$-Galilei algebra. Notice that with the same definition of the time evolution given in \eqref{evolution}, using the deformed coalgebra structures in \eqref{coalgebra poincare}, the same equation is obtained to the first order in $\kappa^{-1}$ when symmetries are given by the $\kappa$-deformation of the Poincar\'e algebra rather than the Galilei algebra.
Notice also that the key ingredient to obtain the decoherence described by eq. \eqref{Lindblad Galilei} is a deformation of time translation symmetries. However, similar decoherence mechanisms might arise in other models of spacetime noncommutativity if time evolution is governed by a Hamiltonian with a deformed coproduct and antipode.

By studying the evolution of the linear entropy
\begin{equation}
    S(t) = 1- \Tr{\rho^2}\,,\label{eq:entropy}
\end{equation}
we can show that the density operator converges asymptotically to the maximally mixed state proportional to the identity, with all the off-diagonal terms being washed out by the evolution.

The time derivative of \eqref{eq:entropy} gives
\begin{eqnarray}
    \dv{t} S &=& -2\Tr{\rho\, \partial_t \rho} = \nonumber \\ &=& \frac{1}{2\kappa}\Tr{\rho\left(\pmb{P}^2 \rho + \rho \pmb{P}^2 - 2\,P_i\rho\,P^i\right)\Big]}\,,
\end{eqnarray}
where we used $\Tr{\rho \comm{P_0}{\rho}}=0$. This result can be rewritten as
\begin{equation}
    \dv{t} S = \frac{1}{\kappa}\sum_i\Tr{O_iO_i^\dagger} \geq 0\,,
\end{equation}
where $O_i\ebd \comm{\rho}{P_i}$. Therefore, we see  that \eqref{Lindblad Galilei} indeed leads to an increase of the entropy (or a decrease of the purity) of the system.

In order to give a more detailed characterization of the nonunitary evolution, we focus on a free nonrelativistic particle, and use $P_0$ as time evolution operator:
\begin{eqnarray}
      \dv{t}\rho_{pq} = \Big[-i \big(E(p)-E(q)\big) -\frac{1}{2\kappa} (\pmb{p} -\pmb{q})^2 \Big]\rho_{pq}(t)\,,\nonumber\\
\end{eqnarray}
where $\rho_{pq}\ebd \mel{E(p),\pmb{p}}{\rho}{E(q),\pmb{q}}$. The solution of this equation gives the time evolution of the density operator 
\begin{equation}
    \rho_{p q}(t) = \rho_{pq}(0)\exp\left[-i t \big(E(p)-E(q)\big) -\frac{t}{2\kappa} (\pmb{p} -\pmb{q})^2 \right]\,.
    \label{solution lindblad}
\end{equation}
The diagonal elements are conserved, while the off-diagonal  elements are characterized by an exponential damping factor, which realises a localization in energy of the system. This indicates a decoherence process in  the superselected \cite{Schlosshauer_2019} basis of momenta.

For this density operator the linear entropy reads
\begin{equation}
    S(t) = 1- \int \dd[3]{p}\dd[3]{q}\abs{\rho_{pq}(0)}^2 \, e^{-\frac{t}{\kappa}(\pmb{p}-\pmb{q})^2}\,.
\end{equation}
Expanding asymptotically around $t \rightarrow \infty$ we find
\begin{equation}
    S(t) \sim 1-\qty(\frac{\pi\,\kappa}{t})^{\frac{3}{2}}\int \dd[3]{p}{\rho_{pp}(0)}^2\,,
\end{equation}
This quantity converges to $1$ as $t\rightarrow \infty$, making the density operator a maximally mixed state.

The decoherence time of this process is given by \eqref{solution lindblad}:
\begin{equation}
    \tau_D = \frac{2\,\kappa}{(\pmb{p} -\pmb{q})^2}\,.
    \label{decoherence time}
\end{equation}
We can rewrite the time evolution of the density operator \eqref{solution lindblad} in terms of the decoherence rate $\Gamma=\tau_D^{-1}$ as
\begin{equation}\label{eq:Gamma}
    \rho_{pq}(t) = \rho_{pq}(0)e^{-i\,t\,\Delta E-\Gamma\,t}\,.
\end{equation}
The decoherence rate is suppressed by the quantum deformation scale $\kappa$ (which, as we mentioned, is taken to be of the order of the Planck energy scale, $\kappa\sim E_P$) and is amplified by increasing values of $\pmb{p} -\pmb{q}$, in such a way that the decoherence process becomes increasingly more efficient as the energies involved approach the Planck scale.

We provide a comparison with other mechanisms of  gravitational decoherence proposed in the literature in \cref{tab deco time}. To allow for such a comparison we reinstate units of $\hbar$, set $\kappa=E_P$ and consider parallel momenta. Notice that this provides an upper bound on the actual decoherence time, since  $\tau_D \leq \tau_D\qty(\pmb{p}\parallel \pmb{q})$. Then  we can write the decoherence time as
\begin{equation}
    \tau_D\qty(\pmb{p}\parallel \pmb{q}) = \frac{\hbar\, E_P}{E_0\qty(\delta\sqrt{E})^2}\end{equation}
where $E=\pmb{p}^2/2m$ and $E_0=m$. 
\begin{table}
    \centering
{\tabulinesep=1.2mm
        \begin{tabu}{|l|l|l}
        \cline{1-2}
       {\small\bf{Physical source of decoherence}} &{\small\bf{Decoherence time}}&  \\ \cline{1-2}
       {\small Stochastic perturbations with noise temperature $\Theta$   \cite{Anastopoulos:2013zya}} & $\frac{\hbar \, }{k_B \Theta}\frac{E_P^2}{E\,\delta E}$ &\\ \cline{1-2}
        {\small Perturbation around flat spacetime \cite{Breuer:2008rh}} & $\frac{\hbar\,E_P}{(\delta E)^2}$ &  \\ \cline{1-2}
        {\small Thermal background of gravitons at temperature $T$ \cite{Blencowe:2012mp}} &$\frac{\hbar}{k_BT}\frac{E_P^2}{(\delta E)^2}$  &  \\ \cline{1-2}
        {\small Fluctuating minimal length \cite{Petruzziello2021}}  &$\frac{\hbar \,E_P^5}{E_0^2\qty(\delta E^2)^2}$  & \\ \cline{1-2}
        Deformation of symmetries (current model)  & $\frac{\hbar\,E_P}{E_0\qty(\delta\sqrt{E})^2}$ &  \\ \cline{1-2}
        \end{tabu}}
    \caption{Comparison of the behaviour of the decoherence times characterizing different gravitational decoherence scenarios, compared to the model studied in this work (last line).     }
    \label{tab deco time}
\end{table}

The decoherence time \eqref{decoherence time} can be used to derive fundamental constraints on the characteristic scales of quantum systems. To do so, we observe that  if a system is in a state with a given energy spread $\delta E$, the state is characterized by a lifetime \cite{Hilgevoord1996}
\begin{equation}
    \tau_c \gtrsim \frac{1}{\delta E}\,.
    \label{lifetime}
\end{equation}
Decoherence cannot be observed if the decoherence time is longer than the lifetime of the state of the system. Thus, for decoherence to be observable, the decoherence time and the lifetime must satisfy the condition $\tau_D \leq \tau_c$. Conversely, if we do not observe decoherence in a given quantum system prepared in a state with energy spread $\delta E$ we have the opposite bound, $\tau_D \geq \tau_c \gtrsim \qty(\delta E)^{-1}$, namely,
\begin{equation}
    \frac{\qty(\delta\, \pmb{p})^2}{\delta E}\lesssim 2\,\kappa\,.
   \label{bound on p}
\end{equation}

For quantum states whose wave function is localized in spatial momentum, $\frac{\abs{\expval{\pmb{p}}}}{\abs{\delta \,\pmb{p}}} \gg 1$ (from which $\frac{\delta \,\pmb{p}^2}{\qty(\delta\, \pmb{p})^2} \sim 1$), we have from \eqref{bound on p} that

\begin{equation}
    m \lesssim \kappa\,.
    \label{bound on m}
\end{equation}
This bound sets a limit on the mass of quantum systems. Specifically, it is not possible to create states highly delocalized in the position for quantum systems with mass greater than the Planck scale $\kappa$. This can be regarded as a more rigorous derivation of an expected feature of quantum-gravitational physics, namely, that the Planck mass constitutes a limit on the mass of quantum mechanical systems, as it is shown from heuristic arguments for instance in \cite{Ng:1993jb}. Notice that had the momentum dependence in \eqref{decoherence time} been different, e.g. linear, we would not have been able to derive such absolute bound. This bound is a genuine implication of quantum spacetime properties. 

Before closing this section let us remark that the limit \eqref{bound on m} should be taken seriously only for elementary quantum systems. For composite systems it has been suggested that the effects of symmetry deformation  might be suppressed by some power of the number of constituents \cite{Hossenfelder:2014ifa, Amelino-Camelia:2011dwc,Amelino-Camelia:2014gga}.

\section{Outlook}

We showed that a Lindblad equation arises naturally as the evolution equation of quantum systems living in a noncommutative spacetime, where the symmetry generators are deformed and described by nontrivial Hopf algebras. This leads to a fundamental decoherence mechanism by which the density operator of a quantum system asymptotically approaches a maximally mixed state. 
While the decoherence mechanism we derived is due to a modified  quantum evolution and not to the presence of an environment, still the evolution equation we found is described by a Lindlad equation, which is usually associated to environmental decoherence.  It would be interesting to understand what type of environment could lead to such a Lindblad equation. This might also provide insights into the physical origin of the time reversal asymmetry emerging in our model, see  eq. \eqref{solution lindblad}, which is usually an imprint of environmental decoherence.

Our analysis shows that there exists a consistent framework modelling putative quantum spacetime effects in which pure quantum states can evolve into mixed states. This fact is particularly suggestive when seen from the perspective of the black hole information paradox that we mentioned in the introduction. Indeed the possibility that the evolution of quantum fields on a black hole background might be not given by ordinary quantum evolution has been dismissed over the years as problematic and lacking links with fundamental theories of quantum gravity \cite{Srednicki:1992bp}. A natural development on this side would be to check whether the nonunitary evolution we derived can be recast in the form of a qubit toy model of black hole evaporation \cite{Avery:2011nb} thus providing the first fully worked out model of nonunitary black hole evolution derived from first principles. 

The fundamental decoherence effect we uncovered might also be instrumental in explaining the quantum-to-classical transition of primordial cosmological perturbations. In fact, the Fourier modes of such perturbations are expected to have Planckian energy at sufficiently early times. At these scales, the decoherence time \eqref{decoherence time} is of the order of the Planck time, thus making the decoherence process very effective.   In this scenario,  one does not need to invoke any of the additional decoherence mechanisms that have been considered in the past, such as  the interaction of the perturbations with external fields.

Besides the possible implications for the foundational issues just discussed, the emergence of fundamental decoherence might open up several windows to experimentally access the  quantum gravity regime.  
Notice that the  decoherence process discussed in this work is not expected to replace the experimentally well-established environmental decoherence \cite{Schlosshauer_2019, Schlosshauer:2022ycx}, and in fact it provides a negligible contribution in typical experimental setups devised to study this. However, it can  become relevant and observable  in specific regimes, as discussed below.

One example of such experimental opportunities concerns tabletop experiments with optomechanical cavities  \cite{Pfister2016, Petruzziello2021}. In these setups, two cavities are realized by means of four mirrors, one of which is movable. Inside the cavities, two atoms or molecules in a harmonic potential are prepared in an entangled state and are hit by an external laser source (realizing a Raman single photon source). The excitations of these sources lead to single photon excitations and the radiation so obtained in the cavities allows to study the vibrational modes of the oscillators. In the presence of an environment, entanglement between the two oscillators is degraded with a rate $\Lambda_{\text{heat}} = \frac{k_B T}{\hbar Q}$ from mechanical heating at temperature $T$; the presence of additional gravitational decoherence mechanisms would increase such a rate to $\Lambda_{\text{heat}}+\Lambda_{\text{grav}}$, where in our case $\Lambda_{\text{grav}}= \frac{\qty(\delta \pmb{p})^2}{2\hbar M_P} $.
Adapting the analysis presented in \cite{Petruzziello2021} to the peculiarities of the fundamental decoherence mechanism we uncovered, we find that experimental parameters that will be achievable in the nearby future, like temperatures $T\sim 10\,\text{mK}$ and quality factors $Q\sim 10^6$ \cite{Pikovski2012,Teufel2011}, will allow to reach Planck-scale sensitivity on the quantum deformation scale, $\kappa \sim 10^{19} \, \text{GeV}$, 
for quantum systems with mass $M\sim 10^{-28}\, \text{Kg}$.

Another physical framework that can allow to test the relativistic version of our model\footnote{As we showed, the relativistic version, based on a quantum deformation of the Poincar\'e algebra rather than the Galilean algebra, produces the same quantum evolution to the first order in $\kappa^{-1}$.} involves modifications of the oscillation rates for cosmological neutrinos. Generic decoherence mechanisms induce modifications on such rates \cite{PhysRevLett.95.160403,Anchordoqui:2005gj, Coelho_2017} that are driven by the exponential damping factor $e^{-\Gamma\,t}$. Such factor starts giving a significant contribution to the oscillation probability roughly when $e^{-\Gamma\,t}\sim e^{-1}$ \cite{PhysRevLett.95.160403}. In our case $\Gamma \sim \kappa^{-1}(\delta p)^2 \sim \kappa^{-1} (\delta m)^2$, see eq.~\eqref{eq:Gamma} and the discussion around it. Taking  $\kappa \sim 10^{19} \, \text{GeV}$ and $(\delta m)^2 \sim 10^{-3} \, \text{eV}^2$ \cite{Coelho_2017}, we find $\Gamma \sim 10^{-40} \, \text{GeV}$. Therefore the $e^{-\Gamma\,t}$ factor starts producing relevant effects on neutrino oscillation when they propagate over cosmological times $t\sim 10^{15} \, s \sim 10^8 \, \text{yr}$.

Such opportunities to look for experimental signatures of the interplay between quantum and gravitational effects are not to be missed given their potential significant impact in the quest for a theory of quantum gravity.

\section*{Acknowledgements}
We acknowledge support from the INFN Iniziativa Specifica  QUAGRAP. This work contributes to the European Union COST Action CA18108 {\it Quantum gravity phenomenology in the multi-messenger approach}. This research was carried out in the frame of Programme STAR Plus, financially supported  by UniNA and Compagnia di San Paolo.

\bibliographystyle{ieeetr}
\bibliography{referencesnoarxiv}

\end{document}